\documentclass[reprint, showpacs, aps, prl]{revtex4-1}
\usepackage{amsmath}   
\usepackage{graphicx}
\usepackage{hyperref}
\usepackage[utf8]{inputenc}
\usepackage[english]{babel}

\usepackage{soul}
\usepackage[usenames,dvipsnames]{color}
\usepackage[normalem]{ulem}

\begin{document}

\title{Cooperative Mercury Motion in the Ionic Conductor Cu$_{2}$HgI$_{4}$}

\author{Damjan Pelc, Igor Markovi\'{c}, and Miroslav Po\v{z}ek}

\affiliation{Department of Physics, Faculty of Science, University of Zagreb, Bijeni\v{c}ka 32, HR-10000, Zagreb, Croatia}

\begin{abstract}

We present the observation of glasslike dynamic correlations of mobile mercury ions in the ionic conductor Cu$_{2}$HgI$_{4}$, detected in both NMR and nonlinear conductivity experiments. The results show that dynamic cooperativity appears in systems seemingly unrelated to glassy and soft arrested materials. A simple kinetic two-component model is proposed, which seems to provide a good description of the cooperative ionic dynamics.

\end{abstract}


\maketitle

In glass-forming materials particles increasingly move together as the glass transition is approached \cite{dyn_het,glass_nonlin}. Such cooperativity is also found in other arrested systems \cite{colgel,gran1,gran2,foam} and seems to be intimately connected to the slow dynamics. Here we report on the observation of large-scale dynamic correlations in a distinctly nonglassy system---the conductive phase of the ionic conductor Cu$_{2}$HgI$_{4}$. Using carefully designed nuclear magnetic resonance experiments we prove that mercury ions are the main contributors to conduction (establishing Cu$_{2}$HgI$_{4}$ as the first known mercury conductor), and show that mercury diffusion is anomalous. These results urge for a more detailed examination of ionic motion. Therefore nonlinear conductivity measurements are used as a probe for dynamical heterogeneity, revealing a characteristic correlation time scale. To explain the cooperativity we propose a simple model related to previous work on glasses \cite{two, CRR2}, with two essential ingredients---disorder and existence of two kinds of particles, slow (copper) and fast (mercury). We compare the results with recent studies of arrested and glass-forming materials \cite{glass_nonlin,colgel}, thus establishing an unexpected connection between seemingly different fields.

Cu$_{2}$HgI$_{4}$ used in experiments was in powder form, synthetized according to standard procedure \cite{proc}. X-ray diffraction at 300 K showed no appreciable contamination with iodides, and all applied techniques (NMR, DSC, conductivity) saw a sharp transition at $T_{c}=344.7$ K, providing further evidence of phase purity. Free induction decays were used to record NMR line shapes, while a recovery sequence was employed in the $^{63}$Cu relaxation measurements. Conductivity was measured in a two-contact cylindrical cell with graphite electrodes, using a low distortion voltage source and lock-in amplifier. At all temperatures, sample resistance was above 1 M$\Omega$. Low frequency (7 Hz) conductivity agrees quantitatively with previously published values \cite{jacs,neutrons}. The third harmonic current $j_{3}$ provided nonlinear conductivity; the heating contribution to $j_{3}$ was estimated to be small due to large sample resistance, and more importantly, uniform over the employed frequency range (linear conductivity is not peaked). Instrumental harmonic distortion effects were also negligible between $\sim 10$ Hz and $\sim 20$ kHz. All measurements were reproducible after several temperature cycles across $T_{c}$.

\begin{figure}[t]
\includegraphics[width=86mm]{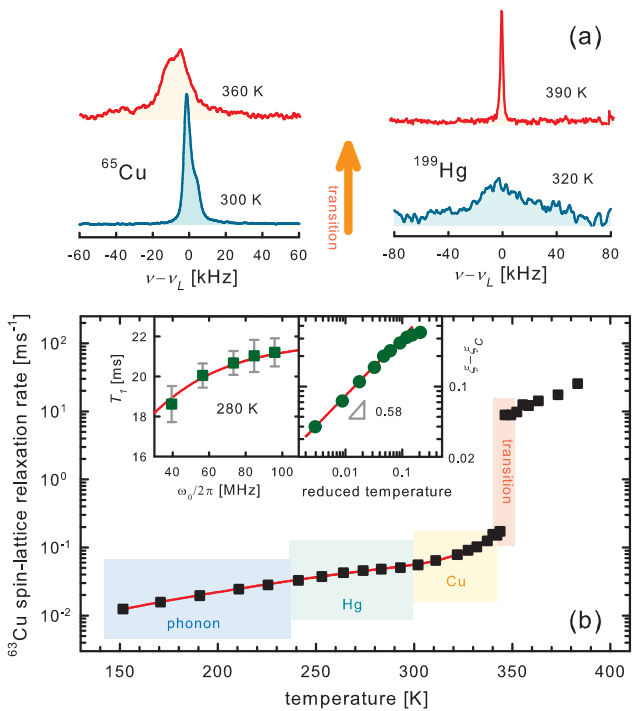}
\caption{\label{NMR} (color online) a) NMR lines of copper and mercury, below and above $T_c$. Frequency is relative to Larmor frequencies, 145.6 MHz for $^{65}$Cu and 91.2 MHz for $^{199}$Hg. Pulse excitation width is $\sim 150$ kHz. Copper lines are normalized to have the same integrals, while the low-temperature mercury line is multiplied by 10 after normalization. Significant motional narrowing is observed for mercury above $T_{c}$. b) Spin-lattice relaxation measurements for $^{63}$Cu. Left insert is frequency dependence at 280 K. Lines are fits obtained from a superposition of phonon and two defect diffusion processes (mercury and copper). Temperature ranges where each of the processes comes into play are indicated. Right insert is order parameter in dependence on reduced temperature.
}
\end{figure}

Before studying cooperative ionic motion in Cu$_{2}$HgI$_{4}$, we must identify the charge carriers and nature of the insulator-conductor transition at $T_{c}$. Ever since the discovery of ionic conduction in Cu$_{2}$HgI$_{4}$ \cite{ketelar}, it has remained unclear which ion species predominantly carries current in the conducting phase above $T_{c}$ \cite{neutrons,merc_struct}. Here we obtain direct proof of mercury motion from NMR experiments---a substantial motional narrowing of the mercury line in the conductive phase [Fig. \ref{NMR}(a)]. In contrast, the copper line is broadened, indicating quasistatic disorder. This suggests that the transition is not melting of the copper sublattice (unlike the related CuI \cite{CuI}), but rather an order-disorder transition \cite{cdil} with the copper ions remaining virtually static. Mercury motion is then enhanced in the changed energy landscape above $T_{c}$. Slow copper diffusion and disorder are essential for explaining mercury dynamic cooperativity, so we perform NMR line shape and relaxation measurements on copper to provide a microscopic picture of the transition.

The structure of Cu$_{2}$HgI$_{4}$ in the ordered phase below $T_{c}$ contains 8 tetrahedral positions per unit cell, with only 3 occupied by copper or mercury \cite{merc_struct}. Thus one expects that a relatively small activation energy is neccessary to create point defects by moving ions from "regular" to normally empty tetrahedral positions. NMR enables us to follow these motions in an ion-specific way, and observe how they lead to a transition to the disordered phase. Naturally abundant copper nuclei are quadrupolar and thus sensitive to electric field gradients (EFGs) present in the material. This is already obvious in the NMR line shapes---the local environment of copper ions doesn't have cubic symmetry, leading to a quadrupolar splitting of the line. While all Cu$^{+}$ which are on regular sites have roughly the same local environment, ions on normally vacant sites (defects) experience much larger EFGs and their NMR lines are substantially shifted and broadened \cite{defects1}. The signal from defects thus contributes as a low, broad background easily separable from the narrow line of "regular" Cu$^{+}$. The integral of the narrow line can be used to obtain the relative number of defects. This number is expected to grow anomalously fast close to an order-disorder transition, as is indeed observed. One can even see power-law behavior close to $T_{c}$,  described reasonably well with a mean-field approach. We define an order parameter $\xi = \left( n_{l}-n_{v} \right) / \left( n_{l}+n_{v} \right)$, with $n_{l}$ and $n_{v}$ the number of regular copper atoms and copper defects, respectively. The transition being first order, $\xi$ drops abruptly to zero above $T_{c}$. A Landau expansion yields $\xi -\xi_{c} \sim \epsilon^{\beta}$ close to $T_{c}$, with $\xi_{c}$ the order parameter at $T_{c}$, $\epsilon = \left(T_{c}-T \right)/T_{c}$ the reduced temperature and $\beta$ the critical exponent: $\beta = 1/2$, comparing well with the experimental value of 0.58 (Fig. \ref{NMR}b right insert). At $T_{c}$ the lattice reorganizes and the distinction between "defects" and "regular" copper ions is lost, as all Cu$^{+}$ positions are equally probable \cite{defects2}. However, no motional narrowing of the copper line is observed above $T_{c}$. Instead, the line broadens due to larger EFGs caused by electrostatic disorder.

To learn more about defect dynamics we measure copper spin-lattice relaxation [Fig. \ref{NMR}(b)]. Diffusing defects create fluctuating EFGs, which influence the relaxation. In addition to a conventional Raman phonon mechanism \cite{abragam}, we observe effects of both mercury and copper defect diffusion on the relaxation rate below $T_{c}$. The contribution from one defect type is \cite{defects1,CuI}
\begin{equation}
\label{T1_dif}
\left.\frac{1}{T_{1}}\right|_{def}  \approx \delta \omega_{Q}^2 n_{v} \left( T \right) \frac{\tau_{h} \left( T \right)}{1+\left[ \omega_{L}\tau_{h}\left( T \right) \right]^{2}}
\end{equation}
with $\delta \omega_{Q}$ a temperature-independent quadrupolar coupling constant, $\tau_{h}$ hopping time and $\omega_{L}$ Larmor frequency. The hopping process is thermally activated \cite{defdif}, with $\tau_{h}=\tau_{0} e^{E_{h}/kT}$, where $1/\tau_{0}$ is the attempt frequency and $E_{h}$ the hopping activation energy. The numbers of defects also follow Arrhenius-type laws except close to $T_{c}$. Combining the temperature and frequency dependences of the relaxation rate, with attempt frequencies estimated from Raman spectroscopy \cite{raman}, we obtain $E_h$ for both copper and mercury defects \cite{suppl1}. The values are $4900 \pm 100$~K and $2220 \pm 50$~K for copper and mercury, respectively. Thus already below $T_{c}$ mercury has a significantly lower hopping activation energy than copper. Microscopic reasons are as of yet unclear.

\begin{figure}[b]
\includegraphics[width=87mm]{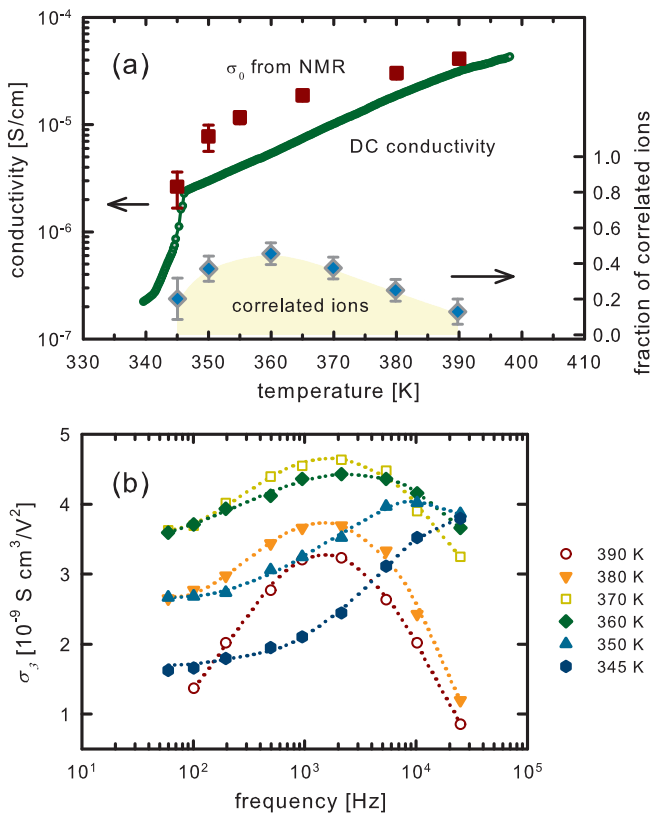}
\caption{\label{corr_panel} (color online) a) DC conductivity (circles) is smaller than conductivity predicted from mercury NMR (squares), indicating anomalous behavior. The discrepancy between conductivities closely follows the number of correlated ions (diamonds, estimated from nonlinear conductivity and NMR) b) nonlinear conductivity in dependence on frequency. Peaks clearly show the existence of a cooperativity time scale. Nonlinear response below $T_c$ is negligible. Lines are guides to the eye. }
\end{figure}
Above $T_{c}$ the mercury diffusion rate increases for an order of magnitude and the $^{199}$Hg line becomes motionally narrowed. The line shape is well fitted by a Lorentzian curve and the hopping time can be extracted from the linewidth using $\Delta \omega = \left( \Delta \omega_{0}\right)^{2} \tau_{h}$, where $\Delta \omega_{0}$ is the static linewidth (below $T_{c}$). The simple formula is valid for $\Delta \omega_{0} \tau_{h} \ll 1$, so we have taken into account corrections for finite $\tau_{h}$ where neccessary \cite{Abragam2}. Employing the Einstein relation for mobility, we can try to calculate the conductivity from extracted hopping times:
\begin{equation}
\label{sigma1}
\sigma_{0} = \frac{2e^{2}n}{kT} \frac{L^{2}}{\tau_{h}},
\end{equation} 
where $2e$ is the charge of Hg$^{2+}$ ions, $n$ their number density ($\sim 5 \times 10^{21}$ cm$^{-3}$) and $L$ a hopping distance of the order of the interatomic spacing ($\sim 1 \text{~\AA}$). If we now take this conductivity and compare it to the measured dc values, we observe the first sign of anomalous behavior: in a region $\sim 30$ K above $T_{c}$, $\sigma_{0}$ is significantly larger than $\sigma_{dc}$ (Fig. \ref{corr_panel}a). Thus relation (\ref{sigma1}), valid for simple stochastic motion of ions, doesn't correctly predict the long-time transport. One may ask if this is due to the existence of some new, intermediate timescale above $\tau_{h} \sim 1$~$\mu$s where motional correlations arise, or just well-known short range correlation effects quantified with the Haven ratio \cite{haven}. To resolve the question, we measure the nonlinear conductivity $\sigma_{3} \left( \omega \right)$, defined by $j=\sigma_1 E + \sigma_3 E^3 + ...$, in dependence on frequency $\omega$ (Fig. \ref{corr_panel}b). Although dynamical correlation effects often bear small influence on linear response, they are intimately related to nonlinear susceptibilities. A quantitative measure is the four-point correlation function \cite{fdt,c4_2,c4_3}, $C_{4} \left( {\bf{y}},t \right) = \langle f \left( {\bf{x}},0 \right) f \left( {\bf{x}}+{\bf{r}},t \right) f \left( {\bf{x+y}},0 \right) f \left( {\bf{x+y+r}},t \right) \rangle_{\bf{x}}$ (with $f$ a suitable dynamic parameter, e.g. intermediate scattering function), representing the correlation of time changes at different points in space. Thus if many ions move synchronously on a characteristic timescale $\tau_{corr}$, $C_{4} \left({\bf{y}},t \right)$ will have a peak at $\tau_{corr}$. Generalized fluctuation-dissipation theorems connect $\chi_{4}$, the spatial integral of $C_{4}$, with the corresponding nonlinear susceptibility \cite{fdt}, making dynamical correlations measurable. In contrast to the case of a dielectric (or magnetic) material, where one measures the response of dipoles to an external field, we detect the response of mobile charges, and the natural response function is $\sigma_3$ instead of the susceptibility $\chi_3$. A lot of activity is currently aimed at modeling dynamical heterogeinity in soft and glassy systems \cite{glass_rev}, but experimental data are still scarce---the first report on nonlinear susceptibility of a glass-former (glycerol) only appeared recently \cite{glass_nonlin}. Here we see similar effects, but in a rather unexpected material. Characteristic correlation timescales are revealed through peaks in $\sigma_{3} \left( \omega \right)$ at frequencies $\omega \tau_{corr} \sim 1$, and the relative number of correlated ions, $N_{corr}$, can be estimated from integrals of the peaks [Fig. \ref{corr_panel}(a)]. In our work we focused on $\sigma_3$ as it is
sensitive to correlated motion; we note, however, that the new correlation timescale also affects the linear conductivity $\sigma_1 \left(\omega\right)$. A shoulderlike feature is visible at the frequency $1/\tau_{corr}$ in the frequency spectrum of $\sigma_1$: this strengthens the analogy with supercooled liquids, where, similarly, the peak of the nonlinear response occurs close to the characteristic relaxation frequency visible in linear response.

\begin{figure}[b]
\includegraphics[width=90mm]{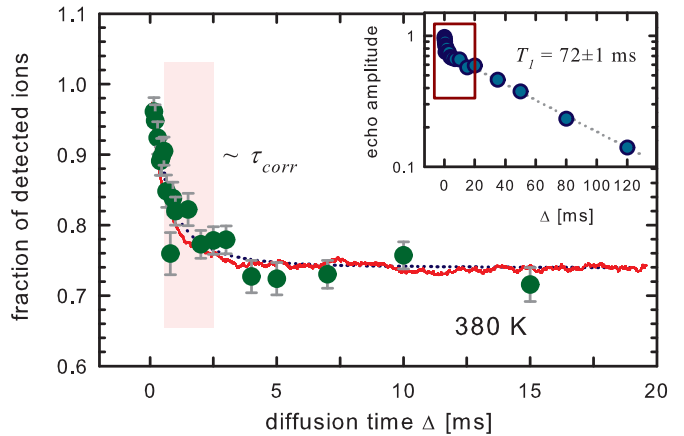}
\caption{\label{SE} (color online) Stimulated spin echo NMR measurements at 380 K, giving evidence of mercury ion trapping at the characteristic timescale $\tau_{corr}$. Insert is raw measurement, the square denoting the zoomed-in segment where dynamic trapping effects are visible. Main graph is compensated for spin-lattice relaxation, showing only the correlation contribution. Full line is from simulation, and the dotted line a stretched exponential fit.}
\end{figure}
Except very close to the transition, $\tau_{corr}$ is substantially longer than the mercury hopping time $\tau_{h}$. Thus a simplistic conduction model can be used to explain the discrepancy between $\sigma_{0}$ and $\sigma_{dc}$. We assume that mercury ions move vigorously most of the time, but sometimes get constrained to small volumes. NMR lines of these ions are broad and do not contribute to the principal narrow line. Occasionally several trapped ions arrange favourably, and leave the "trap" together. Thus the effective number of charge carriers is diminished and the characteristic correlation timescale appears. This is essentially a "cooperatively rearranging regions" (CRR) scenario, well known in glass science \cite{CRR1,CRR2}. A similar mechanism was also proposed for colloidal gel relaxations  \cite{colgel}, and seems to offer a good phenomenological explanation of our data. Contrary to glass-forming liquids, where $N_{corr}$ has no effect on $\sigma_{dc}$, here the correlations influence it. We obtain direct experimental evidence for this model from a different NMR experiment on mercury---stimulated spin echo (SSE) measurements (Fig. \ref{SE}). Moving spins experience much smaller average local fields than trapped ones, leading to a difference in spin decoherence times. This can be exploited to selectively excite and detect only ions moving at a given moment. The SSE sequence \cite{hahn} is perfectly suited for such an experiment \cite{suppl2}. After correcting for spin-lattice relaxation, we can directly observe how, of all ions moving at the time of excitation, a sizeable fraction becomes trapped after a time $\Delta \sim \tau_{corr}$ (Fig. \ref{SE}). This experiment provides an absolute scale for $N_{corr}$, and we can make a comparison with the difference between $\sigma_{0}$ and $\sigma_{dc}$ [Fig. \ref{corr_panel}(a)]. The agreement is gratifying, both in absolute scale and temperature dependence, implying that the difference can be attributed to a diminished effective number of carriers, confirming the phenomenological model.

\begin{figure}[t]
\includegraphics[width=85mm]{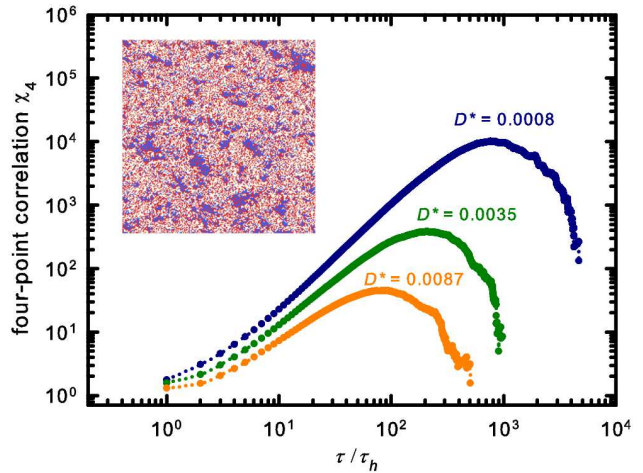}
\caption{\label{siml} (color online) Some results of the 2D simulation. Inserted frame at 2000 steps shows formed mercury islands ($D^{*} = 0.0022$, simulation box 300x300 cells). A correlation timescale is nicely visible in the time dependence of the four-point correlation $\chi_{4}$, for several $D^{*}$. }
\end{figure}

However, microscopic questions remain: what causes confinement, and how are correlated jumps performed? To answer them, we propose a very simple mechanism, related to investigations of spin glasses (essentialy a limiting case of the Edwards-Anderson Hamiltonian with diffusion \cite{c4_1,two}). Aside from disorder, the basic requirement is the existence of two kinds of atoms in the material, with different diffusion coefficients, and short-range interactions. In Cu$_{2}$HgI$_{4}$ this is realized with copper and mercury, on a fixed iodine background. If we take low-temperature activation energies to be representative, we can conclude that mercury diffuses 10$^{3}$ to 10$^{4}$ times faster than copper in the interesting temperature range where correlations appear. As slow copper ions move around, they occasionally form compartments with several trapped mercury ions inside. The compartments can then "open" due to cation rearrangement---once a path is open, many fast mercury ions use it sequentially to empty the compartment. Such behavior is indeed observed in a two-dimensional random walk simulation. Simulations were run on a square lattice with periodic boundary conditions and initially randomly placed ions. In every step the mercury ions moved in random (allowed) directions, while the copper ions moved with a certain probability $D^*$ (which is essentially the ratio of copper and mercury diffusion coefficients). In the course of simulation large mercury "islands" form and dissolve (Fig. \ref{siml}, insert). To calculate the four-point correlation function $\chi_4$, we used the persistence function \cite{ton}, defined as $n_{{\bf{r}}} \left( t_0 \right) = 1$ if nothing has happened on site $\bf{r}$ for $t<t_0$, and $n_{{\bf{r}}} \left( t_0 \right) = 0$ otherwise. $\chi_{4}$ is calculated as the variance of the autocorrelation of $n_{{\bf{r}}} \left( t \right)$, evaluated at all mercury sites \cite{equilibration}. A characteristic correlation timescale is revealed (Fig. \ref{siml}), and the curves qualitatively follow the CRR prediction \cite{ton}. The only parameters we have to set are $D^*$ and effective particle concentration, taking care that the number of vacant sites is above the percolation threshold \cite{note2}. For realistic concentrations and $D^*$, correlation times become about $10^{3} \tau_h$, in fair agreement with experiment. The SSE decay curve (Fig. \ref{SE}) can also be predicted surprisingly well.

The model is "minimal", in the sense that we obtain dynamical heterogeinity with the minimum number of assumptions. This hints at a considerable universality of such correlations. Important effects have, however, been neglected: the iodine lattice potential, short-range electrostatic correlations, electronic dynamics and phonons. Correct temperature behavior cannot be obtained without taking them into account. In contradistinction to glass-forming materials, no slowing down of the correlated dynamics with decreasing temperature is observed in Cu$_{2}$HgI$_{4}$; we suspect that this is due to the additional periodic potential, which tends to restore an ordered state (and succeeds at $T_{c}$). Also, we believe that a full three-dimensional simulation would show similar cooperative behavior (with modified vacancy numbers due to a lower percolation threshold), but this needs to be proven. More elaborate simulations are needed to better understand these issues.

From our results we can conclude that ingredients needed for large-scale motional correlations are quite ubiquitous, so it is reasonable to believe that ionic cooperativity is important for many other systems as well. In disordered ionic conductors it might offer a more convincing explanation of nonlinear response than standard hopping models \cite{nonlin1}, opening up new perspectives for studying ion dynamics. Even more important is the connection with arrested materials, which shows that dynamical correlations are more universal than previously thought. The observed interplay between lattice potential and dynamical heterogeneity is very interesting in itself and could provide a unique possibility for exploring the emergence of glasslike correlations.

\begin{acknowledgements}
We thank D. Cin\v{c}i\'{c} and V. Stilinovi\'{c} for DSC and x-ray measurements and A. Dul\v{c}i\'{c}, H. Buljan, S. Marion and M. S. Grbi\'{c} for helpful discussions and comments. The research leading to these results was supported by equipment financed from the European Community's Seventh Framework Programme (FP7/2007-2013) under Grant Agreement No. 229390 SOLeNeMaR and by funding from the Croatian Ministry of Science, Education and Sports through Grant No. 119-1191458-1022.
\end{acknowledgements}

\clearpage
\section{Supplemental Material}

\section*{Ionic diffusion and copper spin-lattice relaxation} 

\begin{figure}[h]
\includegraphics[width=90mm]{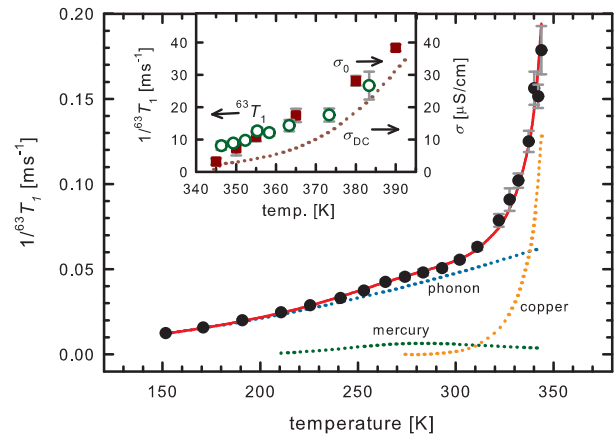}
\caption{\label{T1_lin} $^{63}$Cu spin-lattice relaxation rate below $T_{c}$, on a linear scale. Raman phonon process and two defect diffusion contributions are plotted separately. Insert is relaxation rate above $T_{c}$ (circles), which is seen to roughly fall between $\sigma_{0}$ (squares) and $\sigma_{dc}$ (dotted line).}
\end{figure}

Information on the diffusion of ions and pretransitional increase of copper defect numbers is mainly obtained from line shape and spin-lattice relaxation measurements on copper. The total relaxation rate is a sum of the phonon contribution and the two defect parts: 
\begin{equation}
\label{T1}
1/T_{1} = \left.1/T_{1}\right|_{ph}+\left.1/T_{1}\right|_{Cu-def}+\left.1/T_{1}\right|_{Hg-def}
\end{equation}
Quadruploar spin 3/2 copper nuclei couple to electric field gradients due to phonons and defects; the phonon part is well approximated by $\left.1/T_{1}\right|_{ph} = C T^{2}$ above the Debye temperature (with $C$ a constant) \cite{abragam}. In the simplest approximation of point charges and exponential autocorrelation, the relaxation rate due to one kind of defects is given by Eq. (1) in the paper. Hopping times and defect numbers are temperature-dependent and in principle Arrhenius-like, except copper defects close to the order-disorder transition. A linear scale plot of spin-lattice relaxation vs. temperature below $T_{c}$ (Fig. \ref{T1_lin}) clearly demonstrates that two distinct defect contributions are neccesary to reproduce the temperature dependence of the relaxation. We must therefore fit a superposition of the form of (\ref{T1}), with many adjustable parameters. To constrain the fit, we used several additional results. We easily determined the number of copper defects using measured integrals of the NMR line, as described in the paper. An additional field-dependent relaxation measurement was performed at 280 K to increase the reliability of the remaining fitting parameters---two activation energies for mercury (defect formation and jump) and one for copper (jump), and three multiplicative constants; field and temperature dependecies were fitted simultaneously. Attempt frequencies were constrained to one order of magnitude using Raman spectroscopy data (where anomalous low-frequency modes are identified with the attempt frequencies \cite{raman}), and agree with typical values for similar compounds (e.g. CuI \cite{CuI}).

The activation energies resulting from the fit are given in the paper. The approximation of independent defects is most probably adequate for temperatures not very close to $T_{c}$, but the reliability of our fitting model can be questioned near $T_{c}$, as we suppose that only the number of copper defects grows dramatically. However, the line shape measurements indicate that indeed copper defects are much more prevalent than mercury---intensity of the $^{63}$Cu line drops sharply, while $^{199}$Hg shows no appreciable change (within the relatively large error margin). Above $T_{c}$ the relaxation rate is significantly higher than in the low temperature phase, due to the large number of fast mercury ions. If we assume that the relaxation model of Eq. (1) in the paper is at least approximately valid above $T_{c}$, clearly the contribution of mercury ion diffusion will dominate the relaxation rate (as we know from the absence of motional narrowing in Cu that copper motion is much slower, and $\omega_{0} \tau_{h} \gg 1$). We can thus use mercury hopping times obtained directly from NMR line shapes to make a comparison with the $^{63}$Cu relaxation. In this limit $1/^{63}T_{1}$ is roughly proportional to $n/\tau_{h} \sim \sigma_{0}$ (with $n$ the number of mobile ions, and $\sigma_{0}$ the conductivity as defined in Eq. (2) in the paper). When $1/^{63}T_{1}$, $\sigma_0$ and $\sigma_{dc}$ are plotted on top of each other, we see that the relaxation rate follows the conductivities within the error margin (Fig. \ref{T1} insert), confirming our qualitative analysis.

\section*{Stimulated echo measurements on mercury} 

\begin{figure}[h]
\includegraphics[width=77mm]{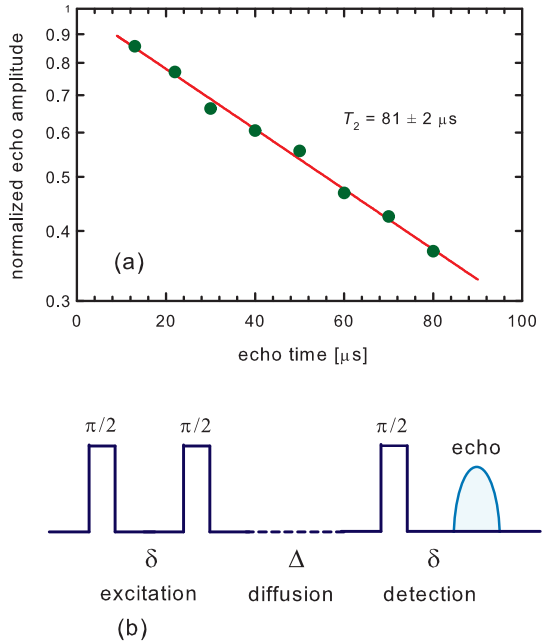}
\caption{\label{stimecho} a) Spin-spin relaxation for mercury at 380 K, obtained using a conventional spin echo sequence.  b) The stimulated spin echo (SSE) sequence. Radio-frequency pulses and spin echo signal are represented schematically.}
\end{figure}

Motional narrowing enables us to make selective NMR measurements on moving mercury ions. The spin-spin relaxation time of stationary ions, $T_{2,trap}$, is of the order of $2 \pi/\Delta \omega_{0} \sim 10$ $\mu$s, while the fast ions have some ten times longer relaxation times (Fig. \ref{stimecho}a). If we use a stimulated spin echo sequence\cite{hahn} (Fig. \ref{stimecho}b) we can exploit this difference. The first two pulses excite only spins which are moving at the moment of excitation (if $T_{2,trap} \ll \delta$ and $\delta \sim T_{2,fast}$) and thus 'tag' them magnetically. After a diffusion time $\Delta$ the third pulse is applied, causing a stimulated echo. 
However, tagged spins which have been trapped during the time $\Delta$ and are stationary at the moment of the application of the third pulse, have shorter $T_{2} \sim T_{2,trap}$ and do not contribute to the echo signal. Thus the echo amplitude should be anomalously small for diffusion times comparable to and larger than the dynamic correlation time $\tau_{corr}$. The size of this 'dip' in diffusion time dependence provides an estimate of the absolute number of correlated ions. Such behavior is observed in simulation as well---even the same functional dependence (a stretched exponential) can be fitted to experimental and simulation results. It is possible to obtain a good agreement between experiment and simulation in both correlation time and fraction of correlated ions (for a vacancy fraction of 0.4 and effective copper diffusion coefficient $D^{*} = 0.0008$), but the simplifying assumption of single copper and mercury jump probabilities makes the stretching parameter of the simulation curve somewhat closer to 1. This could be quickly amended by introducing a jump probability distribution 'by hand', but was deemed physically untransparent. The agreement between simulation and experiment is already quite impressive in the simple version, and more sophisticated models are needed to account for the finer effects. In absence of trapping, the echo decay should be a simple exponential in $\Delta$, with the decay time equal to the spin-lattice relaxation time $T_{1}$. The spin-spin relaxation is single exponential (Fig. \ref{stimecho}a), the line shape doesn't change noticeably for any $\Delta$, and $^{199}$Hg is a spin 1/2 nucleus. Hence the deviations from an exponential decay in the SSE experiment can only be due to dynamical trapping.


\begin{thebibliography}{1}
\bibitem{defects2} This is nicely seen in x-ray diffraction experiments, Ref.~[13].
\bibitem{Abragam2} A. Abragam, (Ref.~[18]). Even at the lowest temperature (345 K) the correction to a Lorentzian result is less than 20$\%$.
\bibitem{equilibration} The system is given an equilibration time of several (typically 10) $\tau_{corr}$, after which the calculation of $\chi_{4}$ begins.
\end{thebibliography}
\end{document}